# A MAPS Detector for High Resolution Low Dose EBSD

High Sensitivity and High Resolution Enable New Analyses

Barnaby D.A. Levin[1], Kalani Moore[1], Nicolò M. Della Ventura[2], McLean P. Echlin[2], Tresa M. Pollock[2], Daniel S. Gianola[2].

1. Direct Electron LP, San Diego, California, USA
2. University of California, Santa Barbara, USA

## Abstract

*The use of highly sensitive pixelated direct detectors has dramatically improved the performance of high energy instrumentation such as transmission electron microscopy. Here, we describe a recently developed monolithic active pixel sensor designed for low energy scanning electron microscopy applications. This detector enables electron backscatter diffraction (EBSD) at lower energies and dose than are accessible with existing scintillator-coupled detectors, expanding grain orientation mapping capabilities to materials such as ceramics that are poor electron conductors. The high detector sensitivity allows collection of rich diffraction information - providing dislocation defect contrast that is otherwise not accessible via EBSD. Indeed, even the energy of single electron interaction events can be measured with this detector, which we demonstrate to energy filter diffraction patterns revealing details of how diffraction occurs at low energy.*

# Introduction

Monolithic Active Pixel Sensor (MAPS) detectors are a form of direct detector introduced to electron microscopy in the 2000s [1]. In cryogenic electron microscopy (cryo-EM) in particular, they have become widespread as an alternative to scintillator-coupled detectors, and have been credited with facilitating a resolution revolution in the field due to their high detective quantum efficiency (DQE) [2]. MAPS detector performance at typical cryo-EM electron beam energies (100 – 300 keV) is limited by the fact that electrons are not stopped by the sensor's sensitive layer, leading to "Landau noise". This can be eliminated using specialized counting algorithms, but these require sparse imaging conditions [3]. However, electron scattering physics indicates that the sensitive layer of a MAPS detector can be sufficiently thick to fully stop an incident electron at lower beam energies (<30 keV) (Fig 1 a-b) [4], allowing for Landau-noise-free, high DQE imaging without the need for a counting algorithm, at significantly higher exposure rates than are possible at high keV. This suggests that MAPS detectors may be well suited to low keV electron microscopy techniques. Previous work has pointed to low energy and photoemission electron microscopy as applications for MAPS detectors [5,6]. Another application is electron backscatter diffraction (EBSD), which is a scanning electron microscopy (SEM) technique widely used for crystallographic mapping. Scintillator-coupled detectors are widely used in EBSD, but their relatively inefficient detection mechanism constrains the application of EBSD to the higher range of SEM accessible energies (>10 keV ) and relatively high beam doses. This limits EBSD's use for studying charge sensitive materials such as poorly conducting ceramics, materials composed of low atomic number elements that weakly scatter, and dose sensitive materials where phase evolution or structural changes can occur such as in magnesium or lithium battery materials. The use of a high DQE MAPS detector may allow EBSD to be used to study these more sensitive specimens. Here we review recent unique EBSD results obtained using the SEMCam MAPS detector.

## The SEMCam

The SEMCam (Direct Electron LP, San Diego, CA, USA) has a 6.5 µm pixel pitch and a 4096 x 4096 pixel full frame size. The detector operates at 92 frames per second (fps) unbinned, and 281 fps with hardware binning at full frame size. Further speed increases can be achieved by reading fewer rows of pixels from the detector, up to ~8000 fps for a readout of 64 rows [7]. The SEMCam used in the work in this article is installed on a Thermo Fisher Scientific Apreo-S SEM (Fig 1 c-d).

## EBSD Results

Across a range of primary SEM beam energies, (4-28 keV), the SEMCam is capable of acquiring electron backscatter diffraction patterns (EBSPs) with richer details than conventional scintillator-coupled detectors [7]. When studying dose-sensitive specimens, beam damage and charge effects are minimized at primary beam energies of 10 keV or below (Fig 2 a). At such energies, the SEMCam can capture EBSPs of higher quality than a conventional detector, even at 10x lower beam doses (Fig 2 b,c). This allows for materials historically unsuited for EBSD to be characterized using the SEMCam, as was recently demonstrated with crystallographic mapping of micro-cracked hafnia, hafnon, and hafnia/hafnon composites [8]. Given that the beam interaction volume in the specimen is smaller at low keV, the high performance of the SEMCam at low keV can also facilitate high spatial resolution mapping.

Recently commercialized hybrid pixel direct detectors for EBSD may also offer some benefits at <10 keV, but these detectors have large (>50 µm) pixels, and consequently have much smaller pixel arrays than the SEMCam. The SEMCam's 4096 x 4096 pixel array allows high-resolution EBSPs to be recorded with rich detail that can be leveraged for novel analysis techniques. One example of how this extra pattern fidelity can be leveraged is to increase the reliability of dictionary indexing (DI) for crystallographic mapping, particularly for low symmetry materials [8]. A 2nd example is the use of EBSP sharpness to image dislocation cells in complex deformed microstructures in bulk specimens, which may have applications in additive manufacturing [9]. A 3rd example is the recently described orientation adaptive virtual aperture (OAVA) technique [10] (Fig 3). This utilizes the concept of virtual apertures applied to diffraction patterns, which are common in 4D STEM [11,12]. In OAVA, virtual apertures on EBSPs are dynamically aligned to local crystallographic orientation to produce images of dislocations, using specific diffraction conditions that are sensitive to the nature of the lattice defects, across a large field of view for single crystal or polycrystalline specimens. A 4th example is to measure the broadening of Kikuchi bands caused by thermal diffuse scattering to calculate local temperature across a specimen [13]. Experimental results suggest temperature can be measured accurately to within ~13 K, and this nanothermometry SEM technique could be applied to non-invasive thermal mapping of microelectronic structures. A final example is strain mapping, which an SEM equipped with a SEMCam may be able to perform at spatial resolutions approaching those of scanning precession diffraction in transmission electron microscopy [14].

In addition to collecting high-resolution EBSPs, the SEMCam can collect energy-resolved EBSD maps. Here, the counting algorithms used with MAPS detectors for cryo-EM can be modified to account for the fact that low-keV electrons are fully stopped in the sensor. This means that under sparse imaging conditions, the energy of each incident electron can be measured. This has recently been used to confirm an energy-related spatial distribution of backscattered electrons [15]. Using a 12 keV primary beam, backscattered electrons with energies as low as 2-8 keV still contributed to Kikuchi patterns, despite having undergone significant inelastic scattering. Filtering the data such that only higher energy (10-12 keV) electrons were included yields sharper diffraction patterns than those obtained from the entire energy range. The ability to perform energy-resolved measurements on EBSPs, to a resolution of ~1 keV, has the potential to allow EBSD to be used as a spectroscopic, as well as a crystallographic tool.

## Summary

Direct sensing EBSD detectors are reshaping the capabilities of diffraction based EBSD techniques in the SEM. This includes expanding this modality for use on material systems that previously were inaccessible, such as charge and dose sensitive ceramics. Energy-resolved EBSD information and the use of energy filtering are helping to better understand the underlying physics of back-scattered diffraction at low energies and shaping what assumptions are built into the simulations that support orientation indexing methods.  The orientation adaptive virtual aperture method we described allows for measurements of dislocation content and character, even for statistically stored dislocations and their networks, which has never been accessible via the EBSD modality before and is a straightforward way to evaluate defects purely via post-processing.

In the future, we expect that the extension of EBSD to light and dose sensitive materials, such as those used for battery and quantum sensing and computing applications will benefit greatly from the newfound ability to map structure and evolution during device charge/discharge cycling. We also expect that direct sensing EBSD information will be extremely valuable for additively manufactured metals, where the microstructure across many length scales is particularly critical for performance.

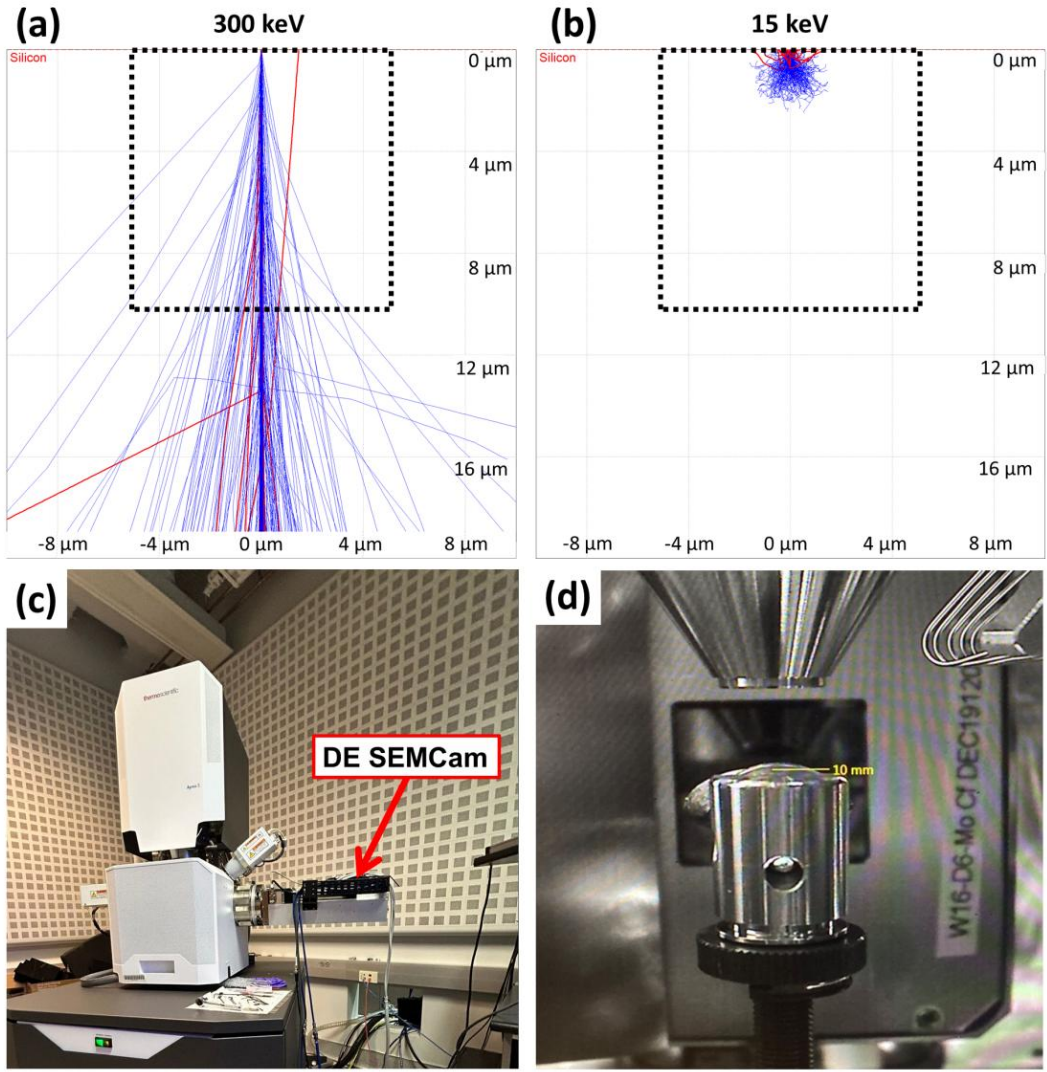

**Fig. 1:** Cross-sections of simulated electron scattering distributions [4] of electrons directly incident on silicon with incident energies of **(a)** 300 keV and **(b)** 15 keV. The dashed box in both plots represents approximate dimensions of a MAPS pixel (~10 μm). **(c)** Photograph SEMCam installed on an SEM at USCB **(d)** Image of the SEMCam detector in the microscope chamber.

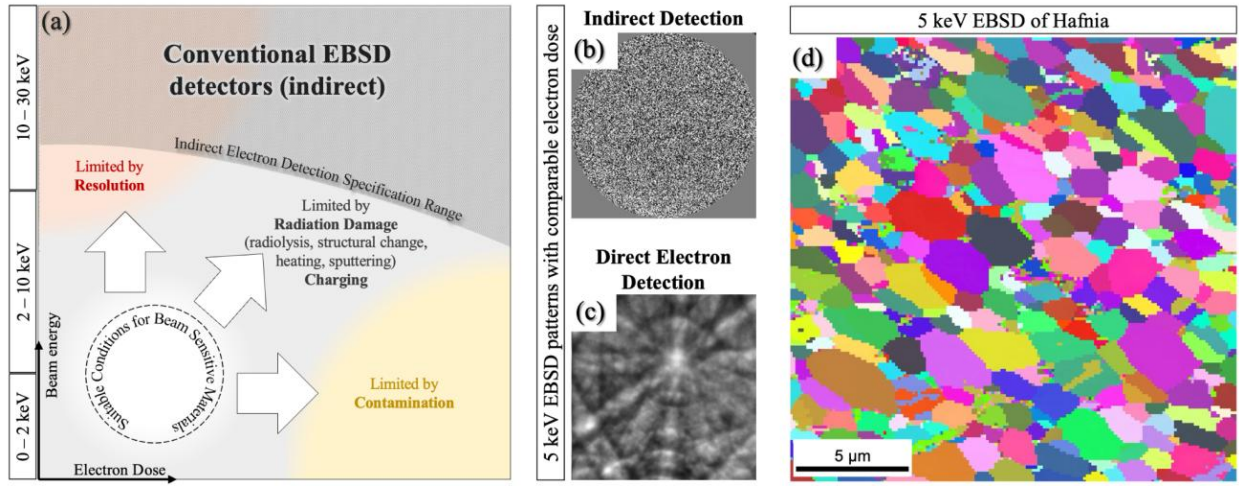

**Fig. 2: (a)** Conventional indirect phosphor-coupled EBSD detectors are limited by resolution, radiation damage and charging, and by contamination effects. **(b,c)** Direct detectors prevail in sensitivity, especially for low probe energies, which has been critical for revealing the microstructure in low crystal symmetry, dose sensitive ceramic materials **(d)**. Figures are adapted from Reference [8] under the CC BY license (http://creativecommons.org/licenses/by/4.0/).

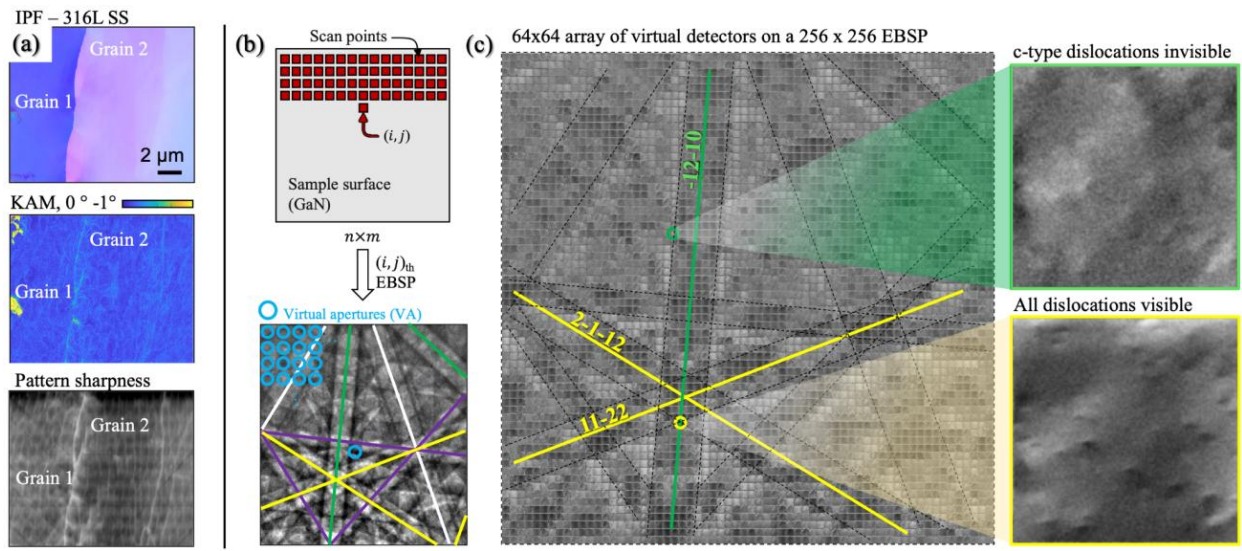

**Fig. 3: (a)** Subgrain microstructural features, such as the cellular structures and low angle grain boundaries that result from the high thermal gradients experienced during additive manufacturing, are composed of dislocation networks that can be revealed using **(b)** orientation adaptive virtual apertures (OAVAs). **(c)** When careful electron diffraction conditions are chosen and applied in post processing to the EBSD datasets, dislocation character and preferential contrast is revealed. (a) is adapted from reference [9] with permission from Elsevier. (b) and (c) are adapted from reference [10] under the CC BY license (http://creativecommons.org/licenses/by/4.0/).

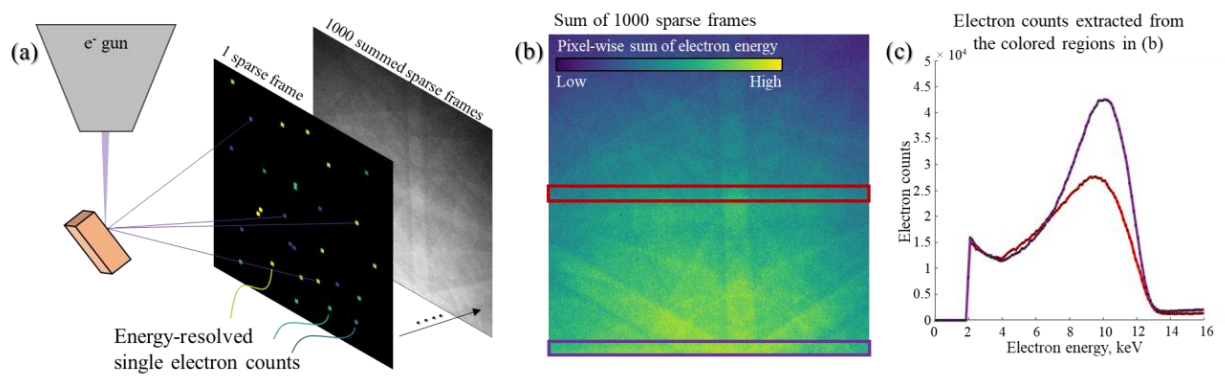

**Fig. 4: (a)** Energy resolved EBSD maps are collected by leveraging electron beam conditions that sparsely illuminate the direct detector, which is calibrated for electron energy and utilizes clustering algorithms developed for TEM. **(b)** The resulting energy-resolved EBSD maps show the spatial distribution of detected electron events, from which cross-sections are extracted at the red and blue boxes plotted in **(c)**.